# Insulator-to-metal transition in sulfur-doped silicon

Mark T. Winkler[1], Daniel Recht[2], Meng-Ju Sher[1], Aurore J. Said[2], Eric Mazur[1,2], Michael J. Aziz[2]
[1]*Department of Physics and* [2]*School of Engineering and Applied Sciences, Harvard University, 9 Oxford Street, Cambridge, Massachusetts 02138*

We observe an insulator-to-metal (I–M) transition in crystalline silicon doped with sulfur to non-equilibrium concentrations using ion implantation followed by pulsed laser melting and rapid resolidification. This I–M transition is due to a dopant known to produce only deep levels at equilibrium concentrations. Temperature-dependent conductivity and Hall effect measurements for temperatures $T > 1.7$ K both indicate that a transition from insulating to metallic conduction occurs at a sulfur concentration between 1.8 and $4.3 \times 10^{20}$ cm$^{-3}$. Conduction in insulating samples is consistent with variable range hopping with a Coulomb gap. The capacity for deep states to effect metallic conduction by delocalization is the only known route to bulk intermediate band photovoltaics in silicon.

Recently, it has been suggested that high efficiency photovoltaic (PV) devices could be fabricated by including an intermediate band of electronic states within the band gap of a traditional PV material [1]. This intermediate band could facilitate absorption of low energy photons and increase photocurrent without reducing cell voltage. An intermediate band photovoltaic (IBPV) device could thus exceed the Shockley-Queisser efficiency limit for single-gap materials [2]. Evidence of this effect has been observed in dilute compound-semiconductor alloys [3] and quantum dot structures [4]. Another proposed IBPV design is a semiconductor doped with impurities that introduce electronic states deep in the band gap [5]. Deep-level impurities have long been considered candidates for absorbing low energy photons [6]; however, they are generally active in non-radiative processes that reduce carrier lifetime and thus reduce PV device efficiency. It has been proposed that these parasitic effects could be avoided if the electrons associated with deep-level impurities delocalize, for example, via a Mott metal-insulator transition [5]. In this scenario, the strong electric fields associated with localized states — ultimately responsible for facilitating non-radiative processes — would vanish. Optical transitions facilitated by deep-levels could then be exploited while avoiding their parasitic effect on carrier lifetime; evidence of this lifetime-recovery effect has been reported for Si doped with high titanium concentrations [7]. However, direct evidence of a delocalization transition in Si doped with deep-level impurities – measured as an insulator-to-metal (I–M) transition – has not previously been observed.

In this Letter we report that an I–M transition occurs in sulfur-doped crystalline silicon. When doping is performed as described below, the transition occurs at sulfur concentrations between 1.8 and $4.3 \times 10^{20}$ cm$^{-3}$. We observe the transition by measuring the temperature-dependent conductivity and Hall effect between 1.7 K and 293 K. For samples doped to a peak sulfur concentration $c_{pk} = (3.6 \pm 0.7) \times 10^{20}$ cm$^{-3}$ the carrier concentration does not change over the temperature range studied, a negative temperature coefficient in the conductivity is observed below 4 K, and conductivity over 100 (Ω·cm)$^{-1}$ persists to $T < 1.7$ K. At the lowest sulfur concentration studied ($c_{pk} = (1.2 \pm 0.3) \times 10^{20}$ cm$^{-3}$), sample conductivity exhibits strong thermal activation, with donor freeze-out and variable-range hopping observed at low temperatures. The remaining samples exhibit some of both sets of behaviors. Conductivity at 2 K varies by a factor $> 10^6$ among samples in this relatively narrow sulfur concentration range.

In all samples, sulfur concentration exceeds the maximum solid solubility of sulfur in Si ($3 \times 10^{16}$ cm$^{-3}$) by a factor of about $10^4$ [8]. We achieve these concentrations using sulfur-ion implantation followed by nanosecond pulsed laser melting (PLM) and rapid resolidification [9]. This method has been demonstrated previously for doping with heavy chalcogens (S, Se, Te) [10, 11]; similar doping concentrations have been achieved via fs-laser techniques [12]. For both techniques, Si doped with about 1% atomic sulfur exhibits strong sub-band gap absorption [10-12], an attractive optical property for IBPV devices.

The optical and electronic properties of Si doped with sulfur to equilibrium concentrations have been reviewed [13] previously. Those experiments found that sulfur introduces discrete deep-level electronic states 100–300 meV below the Si conduction band edge, and that Si is an insulator under these conditions. Because sulfur introduces energy levels far from silicon's band edges, it represents a promising candidate for demonstrating IBPV devices. Density-functional theory calculations have explored the properties non-equilibrium sulfur concentrations in Si [14]; but no experimental studies of electronic transport have been reported yet.

Mott originally described the delocalization of donor electrons in a semiconductor host as an electron-screening effect [15]. At low donor concentrations, the electric field exerted by a donor nucleus on its unpaired electron experiences only dielectric screening. In this regime, all ground-state electrons are localized. When the donor concentration increases above a critical donor concentration $n_{crit}$, metallic screening produced by delocalized electrons eliminates the bound state and donor electrons delocalize. This transition is experimentally observed as an insulator-to-metal (I–M) transition. For a variety of systems [16], $n_{crit}$ approximately satisfies

$$a_b n_{crit}^{1/3} = 0.25, \qquad (1)$$

where $a_b$ is the effective Bohr radius of donor electrons. For shallow levels in a doped semiconductor (such as P or B in Si), the I–M transition has been studied extensively [17]. In this case Bohr radii are on the order of 10 nm and Eq. 1 predicts $n_{crit} \approx 10^{18}$ cm$^{-3}$, in good agreement with measured values [17]. Because the critical concentrations are less than the solubilities [18] of these elements in Si ($>10^{20}$ cm$^{-3}$), traditional local-equilibrium growth techniques can readily provide high-quality metallic and insulating samples. Deep levels, alternatively, have more tightly bound electrons. Thus according to Eq. 1, the I–M transition should occur at higher concentrations ($n_{crit} > 10^{18}$ cm$^{-3}$) than for shallow donors. However, the maximum solubilities of deep-level impurities in Si are generally below $10^{17}$ cm$^{-3}$ [18]. Accordingly, equilibrium doping does not lead to an I–M transition for these elements in Si. This work is the first demonstration of an I–M transition in crystalline Si driven by a deep-level dopant.

Single-crystal Si wafers (boron doped, $\rho \approx 25$ Ω·cm) were commercially ion-implanted, nominally with 95 keV $^{32}$S$^+$ to doses of (3, 7, 9, and 10)×10$^{15}$ cm$^{-2}$. The implanted region, amorphised by the implant, was melted using four spatially homogenized XeCl$^+$ excimer laser pulses (fluence = 1.7 J cm$^{-2}$, $\lambda$ = 308 nm, pulse duration 25 ns full-width at half-maximum) in laboratory ambient. Using this process the melted region resolidifies as a single crystal free of extended defects, doped with about 1% atomic sulfur [10, 11]. We quantified the sulfur concentration-depth profile using secondary ion mass spectrometry (SIMS), identifying the peak concentration $c_{pk}$ and the retained areal sulfur dose $\Phi$. The sample preparation and characterization process have been described in detail previously [10, 11]. We report on four samples here; we label them A, B, C, and D and outline their properties in **Table 1**.

After the PLM process, samples were cleaned with acetone, methanol, and isopropanol and etched (60 s) in hydrofluoric acid (10%) to remove the surface oxide. We defined cloverleaf test structures by masking samples with photoresist and etching to a depth of 1 $\mu$m using an $SF_6$-based reactive ion etch; the doped region is isolated from the substrate by the rectifying junction between the two [11, 19]. The cloverleaf structures have a total width of 2 mm and central device diameter of 100 $\mu$m (Fig. 2 inset). We deposited Ti–Ni–Ag (20–20–200 nm, Ti adjacent to Si) contacts of 100-$\mu$m diameter at the outer edges of each cloverleaf. Samples were affixed and wire-bonded to non-magnetic chip carriers and mounted in a He cryostat. We measured sample sheet conductivity $\sigma_s$ over the temperature range 1.7–40 K using the van der Pauw technique [20]. The DC excitation current $I$ was selected for each sample to yield 5 fW of resistive heating; self-heating effects were not observed. Hall measurements were performed at 2 K and 293 K using standard techniques [21] and a magnetic field $B = 0.6$ T. Sheet carrier concentration $n_s$ was calculated from the measured Hall voltage $V_H$ using $n_s = r_H IB(eV_H)^{-1}$, where $e$ is the elementary charge and $r_H$ is the Hall scattering factor. We assumed $r_H = 1$, an assumption that is generally accurate in heavily doped Si [22].

Using the values of $c_{pk}$ and $\Phi$ determined from sulfur concentration-depth profiles, we defined an effective doped-layer thickness $d_{eff} \equiv \Phi / c_{pk}$ for each sample. Using this quantity, we calculated the conductivity $\sigma = \sigma_s / d_{eff}$ and carrier concentration $n = n_s / d_{eff}$ from the corresponding sheet quantities. We discuss this approach's accuracy below and argue that it sets lower bounds on the peak values of $\sigma$ and $n$ in the sulfur-doped region.

Figure 1 shows the low temperature conductivity for all samples. At 2 K, conductivity differs by a factor $>10^6$ among samples whose peak sulfur concentration varies by a factor of three. Sample A exhibits a slightly negative temperature coefficient between 2–4 K (inset Fig. 1). In Fig. 2, the carrier concentration at 2 K is plotted against the same value at room temperature. For samples with peak sulfur concentrations of at least $(3.6 \pm 0.7) \times 10^{20}$ cm$^{-3}$, the low- and high-temperature carrier concentration are indistinguishable. For samples with lower sulfur concentrations, the carrier concentration at 2 K is significantly smaller than that at 293 K.

The metallic state is defined by finite conductivity as $T \to 0$, whereas insulators exhibit conductivity that must be thermally activated. As shown in Fig. 1, samples C and D exhibit strongly thermally-activated conductivity and are clearly insulators. Samples A and B exhibit conductivities that vary only slightly over the measured temperature range, and appear to remain finite as $T \to 0$. Below 4 K, sample A exhibits a slightly negative temperature coefficient, as expected for a metal. This effect is small, however, and could result from using an effective layer thickness. The conductivity of sample B increases with temperature (by about 10% from 2 to 10 K); although not typical of a metal, a positive temperature coefficient has been observed in just-metallic semiconductors [23]. Finally, the magnitude of the conductivity at 2 K for samples A and B is relatively large. In previous measurements of Si doped with shallow donors at just-metallic concentrations [23], the conductivity at 2 K is much lower (about 10 $(\Omega \cdot cm)^{-1}$) than the conductivity we observe. We thus conclude that sample A is metallic, whereas samples C and D are insulating; sample B appears to be near the transition point. According to these data, the transition between the insulating and metallic states in sulfur-doped Si occurs at peak sulfur concentrations between $(2.2 \pm 0.4)$ and $(3.6 \pm 0.7) \times 10^{20}$ cm$^{-3}$.

The Hall effect data of Fig. 2 are consistent with this conclusion. As expected for a doped semiconductor in the metallic state [24], the carrier concentrations of samples A and B are constant as temperature decreases from 293 to 2 K. In contrast, the carrier concentrations of samples C and D are substantially smaller at 2 K than at room temperature — consistent with donor electrons relaxing from thermally-excited conduction-band states into localized ground states as temperature decreases.

Although we have identified the I–M transition with the peak sulfur concentration, $n_{crit}$ likely depends on the microscopic sulfur configuration(s) and the electronic states they introduce. Thus, the critical concentration we report is likely an upper bound on $n_{crit}$ for a particular sulfur defect in Si. For example, metallic samples A and B exhibit room temperature sheet carrier concentrations lower than sulfur dose by about an order of magnitude. It is unclear whether this results from a portion of states remaining localized, or whether a large fraction of delocalized states resides far from the Fermi level and thus do not participate in conduction. Because the distribution of defect states depends, in general, on a material's exact thermal history, it may be challenging to precisely identify $n_{crit}$ for I–M transitions realized via non-equilibrium doping.

At low temperature, the conductivity of a doped semiconductor in the insulating state scales as

$$\sigma(T) = \sigma_0 \exp[-(T_0/T)^s]. \qquad (2)$$

The value of the constant $s$ depends on the temperature and density of states at the Fermi level; the prefactor $\sigma_0$ and exponential activation $T_0$ are related to material parameters by different relationships for each value of $s$ [25]. We fit the conductivity of the insulating samples to Eq. 2 using several different exponents: $s = 1/4$, $1/2$, and $1$, corresponding to Mott's variable range hopping (VRH), VRH with a Coulomb gap, and nearest-neighbor hopping, respectively [25]. Both $s = 1/4$ and $s = 1/2$ provide reasonable fits, with average relative mean square errors of 1.7% and 1.0%, respectively, and fitting ranges restricted to $T < 20$ K and $T < 15$ K, respectively. Acceptable fits cannot be found using $s = 1$. The data and fits for $s = 1/2$ are shown in Fig. 3, with fitting parameters provided in Table 1. To determine the value of $s$, we replotted the data as $W = d\log(\sigma)/d\log(T)$ versus temperature on a log-log scale (inset Fig. 3). For conductivity activated as in Eq. 2, the slope of $\log W$ versus $\log T$ yields the value of $-s$ [25]. By this analysis, sample D yields $s = 0.43 \pm 0.06$ — very close to $s = 1/2$ — indicating that conduction likely occurs by variable range hopping with a Coulomb gap in this sample [25]. The data for sample C cannot be identified with a specific conduction mechanism; regardless, sample C exhibits weaker temperature activation than sample D. In hopping conduction, this behavior is consistent with an increased electron correlation length [25], which would be expected as the dopant concentration approaches $n_{crit}$.

Finally, we comment on our calculation of $\sigma$. Both $c$ and $\sigma$ vary in the sulfur-doped region as a function of distance $z$ from the sample surface. The peak values can be related using $\sigma_{pk} \equiv \alpha c_{pk}$, where $\alpha$ is a constant with dimensions of cm$^2 \cdot \Omega^{-1}$. Because the samples are doped at concentrations near the I–M transition, conductivity rises much more quickly than linearly with sulfur concentration; thus $\sigma(z) \leq \alpha c(z)$, with the equality realized only at the depth of $c_{pk}$. Together with the definition of sheet conductivity $\sigma_s = \int \sigma(z)\,dz$, we can state that $\sigma_s \leq \alpha \int c(z)\,dz = \alpha \Phi$. Using the definitions $\Phi \equiv d_{eff} c_{pk}$ and $\sigma_{pk} \equiv \alpha c_{pk}$, we obtain $\sigma_s \leq d_{eff} \sigma_{pk}$. Thus $\sigma_{pk} \geq \sigma_s / d_{eff}$, and $\sigma = \sigma_s / d_{eff}$ represents a lower bound on the peak conductivity in the implanted region. Above, we emphasized that samples A and B exhibit values of $\sigma$ larger than those exhibited by just-metallic Si doped with shallow donors to support our argument that samples A and B are metallic. Underestimation of $\sigma$ strengthens this argument.

In conclusion, we observe an insulator-to-metal transition in Si doped with sulfur via ion implantation followed by pulsed-laser melting. Conductivity and Hall effect data indicate that the transition occurs at a peak sulfur concentration between 1.8 and $4.3 \times 10^{20}$ cm$^{-3}$. At sulfur concentrations just below the transition, variable-range hopping with a Coulomb gap is observed along with a decrease in the conductivity at $T = 1.7$ K by a factor $>10^6$ relative to metallic samples. This result is the first observation of an I–M transition driven by a deep-level impurity in crystalline Si. The capacity for deep states to effect metallic conduction by delocalization is the only known route to bulk intermediate band devices, including photovoltaics, in materials such as Si for which the carrier lifetime is limited by non-radiative recombination.

Several people contributed to this work. A.J.S. and D.R. fabricated the samples; D.R. also analyzed the SIMS data. M.W. conceived the experiment, performed electronic measurements with M.S., and prepared the manuscript. E.M. and M.J.A. supervised the research. All authors participated in manuscript editing. The authors thank Jacob Krich (Harvard) and Jeff Warrender (U.S. Army ARDEC-Benét) for helpful conversations; Tom Mates (UCSB) for SIMS measurements supported by NSF contract DMR 04-20415; Ed Likovich and Prof. Venky Narayanamurti for access to their cryostat; and Heyun Yin (Varian Semiconductor) for providing S-implanted Si. The research of D.R., A.J.S., and M.J.A. was supported by the U.S. Army ARDEC (contract W15QKN-07-P-0092). The research of M.W., M.S., and E.M. was supported by NSF contracts CBET-0754227 and DMR-0934480. M.W. was supported by an NSF Graduate Research Fellowship. D.R. was supported by the Department of Defense's National Defense Science and Engineering Graduate Fellowship Program. A.J.S. was supported by the Fulbright Fellowship program.

| Sample | $\Phi$ $(10^{15}\,\text{cm}^{-2})$ | $c_{pk}$ $(10^{20}\,\text{cm}^3)$ | $d_{eff}$ (nm) | $T_0$ (K) | $\sigma_0$ $(\Omega\cdot\text{cm})^{-1}$ |
|---|---|---|---|---|---|
| A | $9 \pm 2$ | $3.6 \pm 0.7$ | 250 | – | – |
| B | $10 \pm 2$ | $3.8 \pm 0.8$ | 260 | – | – |
| C | $4.3 \pm 0.9$ | $2.2 \pm 0.4$ | 250 | 9.05 | 3.20 |
| D | $3.0 \pm 0.6$ | $1.2 \pm 0.2$ | 260 | 326.5 | 2.46 |

**TABLE I**. Properties of samples studied in this work, including: sulfur dose $\Phi$, peak sulfur concentration $c_{pk}$, effective layer depth $d_{eff} = \Phi / c_{pk}$, and fitting parameters $T_0$ and $\sigma_0$ defined in Eq. 2.

# Figure Captions

**FIG. 1.** Conductivity of S-doped Si for temperatures 1.7 K < $T$ < 40 K. Sample properties are given in Table 1. Inset, sample A exhibits a negative temperature coefficient.

**FIG. 2.** Temperature-dependence of carrier concentration; dashed line shows metallic behavior. Samples A and B behave as metals, while C and D behave as insulators.

**FIG. 3**. Conductivity of insulating samples fit to Eq. 2 with $s = 1/2$ (variable range hopping with a Coulomb gap). Inset: slope of $W = d\log(\sigma)/d\log(T)$ versus $T$ on a log-log scale identifies the value of $s$. The fit for sample D yields $s = 0.43 \pm 0.06$; solid line shows $s = 1/2$ for reference.

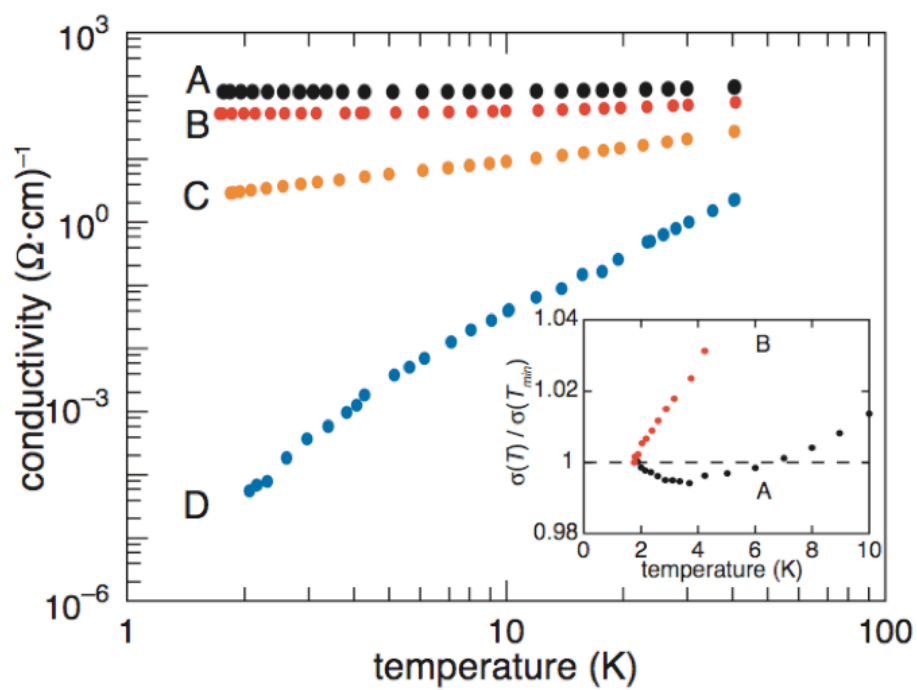

Figure 1 — MT Winkler, D Recht, M Sher, A Said, E Mazur, MJ Aziz

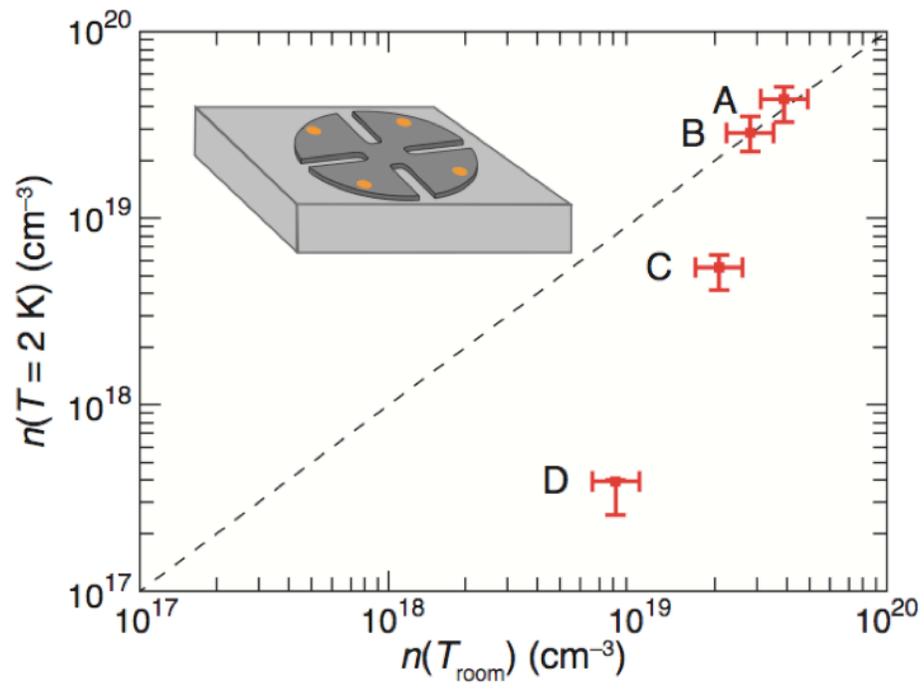

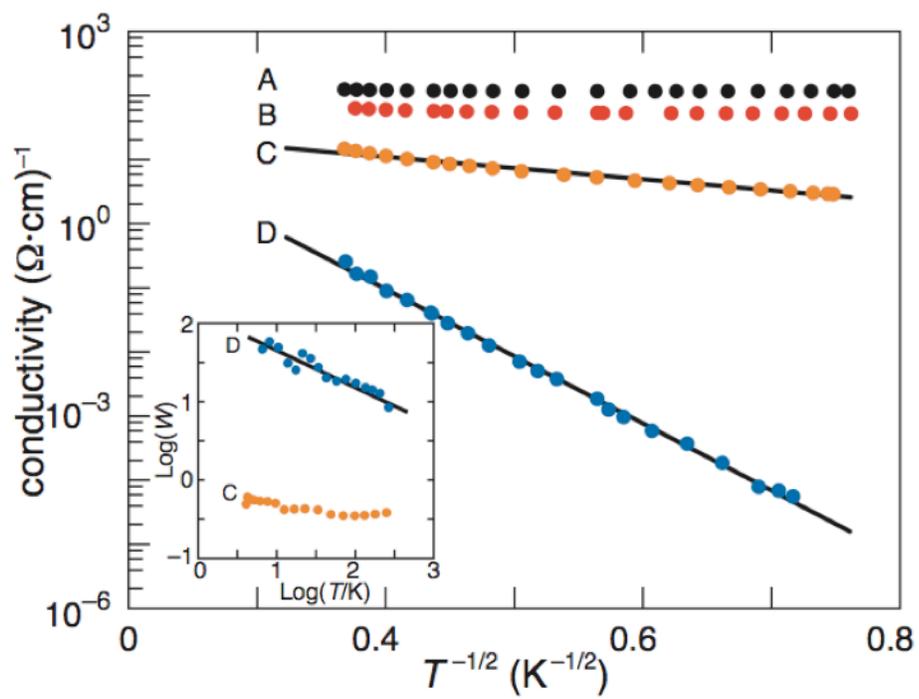

Figure 3 — MT Winkler, D Recht, M Sher, A Said, E Mazur, MJ Aziz